\documentclass{aa}
\usepackage{graphicx}

\begin{document}

\title{Long-term remnant evolution of compact binary mergers}

\author{W. Domainko$^1$ and M. Ruffert$^2$}

\institute{${^1}$Institut f\"ur Astrophysik, Leopold-Franzens Universit\"at
Innsbruck, Technikerstra\ss e 25, A-6020 Innsbruck, Austria\\
${^2}$ School of Mathematics, University of
Edinburgh, Edinburgh EH9 3JZ, Scotland, UK}

\offprints{\email{wilfried.domainko@uibk.ac.at}}

\authorrunning{W. Domainko \& M. Ruffert}
\titlerunning{Remnants of short GRBs}
\date{Received / Accepted}
 
\abstract{We investigate the long-term evolution and observability 
of remnants originating from the merger
of compact binary systems and discuss the differences to supernova remnants.
Compact binary mergers expel much smaller amounts of mass at much higher
velocities, as compared to supernovae and therefore the free expansion phase
of the remnant will be short ($\sim$ 1 - 10 yr).
But merger events with high mass ejection exploding in a low density environment
remain in this ejecta dominated stage for several hundred years and 
in general the remnants will be
observable for a considerable time ($\sim$ 10$^6$ - 10$^7$ yr). 
Events releasing large amounts of 
kinetic energy
may be responsible for a subsample of observed giant HI holes of unknown 
origin as 
compact binaries merge far away
from star forming regions. If the ejecta consist primarily of
actinides, on long timescales the expelled material will contain mainly
the few quasi-stable nuclei in the actinides range. Consequently the 
abundance of each isotope in the ejecta might be of the order of a few percent.
During their decay some actinides will produce observational signatures 
in form of gamma ray lines.
We particularly investigate the gamma ray 
emission of Am 243, Cm 247, Cm 248 and Bi 208 and estimate their 
observability in 
nearby remnants. Detections of the gamma ray lines with INTEGRAL 
will be
possible only in very advantageous cases but these remnants are promising 
targets for future instruments using focusing optics for soft gamma rays.
Due to the low mass expelled in mergers and due to the lack of free
electrons in the ejecta, the merger remnants might be significantly fainter
in bremsstrahlung and synchrotron radiation than comparable supernova remnants.
Hence merger remnants might represent a candidate for very recently discovered
'dark accelerators' which are hard gamma ray sources with no apparent emission
in other bands.

\keywords{Stars: neutron -- binaries: close -- ISM: bubbles -- Gamma rays: 
bursts -- nucleosynthesis -- abundances}
}

\maketitle


\section{Introduction}

Observationally several neutron star binaries are identified to date 
(Stairs \cite{stairs04})
and it is expected that they inspiral as a result of gravitational waves and 
merge subsequently. These 
neutron star neutron star (NSNS) mergers were proposed as central engines of
cosmological gamma-ray bursts (GRBs) already two decades ago (e.g. Paczy\'nski
\cite{paczynski86}) and are nowadays considered as the most promising
explanation for the subclass of short, hard GRBs discovered with BATSE
(Kouveliotou et al. \cite{kouveliotou93}). This scenario was very recently
further supported by the first localization of a short GRB afterglow
(Bloom et al \cite{bloom05},
Gehrels et al. \cite{gehrels05}). Theoretical progress on the physics of
NSNS mergers was made by extensive numerical simulations (Ruffert et al.
\cite{ruffert96}, \cite{ruffert97}, Lee et al. \cite{lee99} and 
Rosswog \& Davies \cite{rosswog02}).

In this letter 
we focus on the impact of the ejecta of such merger events on the surrounding 
medium and on the observational consequences of remnants which are left 
behind these explosions. 
Simulations by above mentioned authors revealed that the mass ejected
in NSNS mergers is typically small (about 10$^{-4}$ -- 10$^{-2}$ M$_{\odot}$),
fast (a significant fraction of the speed of light) and
furthermore that it possibly consists of heavy r-process nuclei. 
The mass of the ejecta
might be increased by neutrino driven outflows of material from the  
quasi-stable super-massive neutron star (lifetime $\sim$ 100 ms) formed in the
merger
before it collapses to a black hole (Shibata et al. 
\cite{shibata05}). 
Additionally to the previously mentioned merger event of two NS
we explore the development of remnants 
which contain a NS and a stellar mass black hole.
Very recently simulations of mergers of neutron star
black hole (NSBH) binaries with small mass ratios were performed by Rosswog 
(\cite{rosswog05}). These simulations showed that NSBH mergers
are unlikely to drive GRBs but  
are likely to produce electromagnetic transients
or even dark explosions,
and that the mass expelled during the merger is generally
much larger than in the NSNS merger case (about 0.01 - 0.2 M$_{\odot}$).

Remnants of GRBs with an hypernovae as central engines have been studied in 
the past with respect to their aspherical shape
(Ayal \& Piran \cite{ayal01}), their capability to produce giant HI
holes in the interstellar medium (Efremov et al.
\cite{efremov98} and Loeb \& Perna \cite{loeb98}) and the lack of
chemical signature left behind (Perna \& Raymond \cite{perna00}). HI holes
might be promising candidates for GRB remnants as some of them 
seem to have no obvious origin (Rhode et al. \cite{rhode99}, Simpson 
et al.
\cite{simpson05} and Hatzidimitrou et al. \cite{hatzidimitrou05}).
For the case of mergers a possible evolution szenario of the ejecta for 
the first few days after the event was discussed
by Li \& Paczy\'nski (\cite{li98}).

In contrast to previous work we explore in this letter the 
long-term evolution of
remnants from compact binary mergers and investigate their
appearance due to the signature of their chemical composition and
due to various other emission mechanisms of the ejecta.


\section{Evolution of the remnants}

We investigate the evolution of
remnants originating from the merger of compact binary systems and also discuss
the difference to supernova remnants (SNRs). 
Compact binary mergers expel
smaller amounts of mass at much higher velocities, as compared to
supernova explosions. We use this fact and the long evolution 
time of the progenitors of mergers for the considerations in
this section. Remnants of very energetic explosions within a surrounding 
medium evolve in several stages (e.g. Chevalier \cite{chevalier77}):

\subsection{Free expansion phase}

The early
expansion of the remnant will be driven mainly by the kinetic energy of the
expelled material. This phase of free expansion will last until the swept up 
mass of the
surrounding medium will equal the mass of the ejecta. Since the ejecta mass
in compact binary mergers is small and the velocity of the material is
very high in comparison to the mass and velocity of material
expelled in supernova explosions, the remnant will reach the stage 
of equal mass very quickly. 
For a NSNS merger expelling only $10^{-3}$ M$_{\odot}$ in a 
medium of density 1 cm$^{-3}$, equality of mass will be reached
within less then 5 years.
However, in the case of
a NSBH merger with a relatively high mass loss of 0.2 M$_{\odot}$ 
(still one order of magnitude lower than the mass ejected in a typical
SN Ia) exploding in a low density environment of $10^{-6}$ cm$^{-3}$ 
typical for the inter-galactic medium (IGM), 
the free 
expansion phase can last for about 1000 years. 
This scenario of NSBH mergers exploding into the IGM is likely as 
compact binaries might be expelled from
galaxies by natal kicks during their formation.
For constraints on the observability of this ejecta dominated phase see
Sect. \ref{brems} and \ref{sync}.

\subsection{Sedov -- Taylor phase}

After the remnant has swept up an amount of mass from the ambient medium which 
is comparable
to the mass ejected during the explosion, the interior of the bubble 
thermalizes and the expansion of the remnant is mainly
pressure driven. This is described by the Sedov - Taylor solution. Due to the 
lower mass of material which is ejected in compact binary mergers, 
the time scale is shorter to reach the pressure driven phase for 
merger remnants than for a SNR. Hence remnants
of mergers will start their pressure driven expansion with a much higher
velocity than SNR. This is of particular interest in a hot medium 
with high sound velocity, as
it has been shown that in hot media the pressure driven evolution of
SNRs deviates from the Sedov - Taylor solution (Dorfi \& V\"olk
\cite{dorfi96}, Tang \& Wang \cite{tang05}). In contrast to SNRs,
the remnants of mergers will enter the pressure driven phase
with a high expansion velocity and the explosion can still be described
with the Sedov - Taylor solution. The evolution of the remnant will deviate
from this approximation only after the expansion has decelerated to a velocity
which is comparable to the sound velocity of the hot ambient medium.
This evolution scenario of remnants is relevant for explosions which happen in
elliptical galaxies containing hot gas or in case of the progenitor being part
of the intra-cluster stellar population and exploding in the hot intra-cluster
medium (ICM) (note that GRB 050509b possibly exploded in the ICM of a 
galaxy cluster, see Bloom et al. \cite{bloom05} and Gehrels et al.
\cite{gehrels05}).

\subsection{Late evolution of the remnant}

The late phase evolution of a blast wave exploding into a cold uniform ambient 
medium will result in a cool thin expanding HI shell (Chevalier
\cite{chevalier74}, and for application on GRB remnants see Efremov et al.
\cite{efremov98} and Loeb \& Perna \cite{loeb98}). The size and expansion
velocity of the HI shells will be constrained by the amount of kinetic
energy released in the explosion which
for compact binary mergers is related to the mass
of material ejected during the event. For NSNS mergers ejecting mass of
$10^{-4}$ $M_{\odot}$ with a velocity of 0.5 c (with c being the speed of
light), the kinetic energy released could be as small as $\sim 2\times 10^{49}$
erg. This is nearly two orders of magnitude smaller than the average kinetic
energy released in a supernova. On the other hand a NSBH merger
ejecting 0.2 $M_{\odot}$ with a velocity of 0.5 c, the kinetic energy would be
of the order of $\sim 5\times 10^{52}$ erg, several times
the kinetic energy produced in an average supernova explosion. Observationally 
identifying late stages of compact binary merger remnants will help  
constrain the energetics of such events. Events with
high mass ejection may even produce giant expanding HI shells. We note
that several large HI rings of unknown origin are
indeed observed in a number of galaxies (Rhode et al. \cite{rhode99}, Simpson 
et al.
\cite{simpson05} and Hatzidimitrou et al. \cite{hatzidimitrou05}). Some of these
observed HI holes do not show young star clusters at their centers which makes
their emergence from multiple supernova explosions quite unlikely.
NSBH mergers expelling a large amount of material might be the mechanism
responsible for some of the observed HI shells as 
compact binaries
take some time to merge during which they move away from
star forming regions. HI rings might even help to identify the sites of
NSBH merges which resulted in dark explosions (Rosswog \cite{rosswog05}).


\section{Observational signatures of the remnants}

In this section we explore various emission mechanisms of the ejecta of compact
binary mergers which consist of possibly
very heavy elements, as compared to supernova ejecta.
Considerations in this section are less well founded as the nature of the 
ejecta in
compact binary mergers is not known very well. We also investigate
the exciting possibility that these remnants might be related to 
the very recently discovered 'dark accelerators'.

\subsection{Signature of r-process elements}\label{rproc}

The ejecta of compact binary mergers consists of exceptionally neutron rich 
material. This might result in the production and distribution of 
heavy r-process elements (Lattimer \& Schramm \cite{lattimer74}, Ruffert et al.
\cite{ruffert97}, Freiburghaus et al. \cite{freiburghaus99}). The exact
composition of the ejected material is not very well understood as
many complicated physical processes play a role in neutron rich
nucleosynthesis (e.g. Goriely et al. \cite{goriely04}). 
For low values of
the relative electron number density of the ejecta, 
nucleosynthesis will lead to the production
of mainly actinides in the expelled material (Ruffert et al. \cite{ruffert97}).
The early composition of the ejecta is difficult to determine
but there might be certain constraints on the composition after a few thousand 
years. The actinides consist of only 18 nuclei with a half live
time longer than 5000 years. Hence we expect that on timescales of a few 
thousand
years the ejecta will be overabundant in the quasi stable nuclei as many of the
decay chains of non stable nuclei will end at the quasi
stable nuclei. So on long timescales if the ejecta consists of less then 20
isotopes we expect, as a first crude approximation, 
an abundance of the order of a few 
percent of every quasi stable specific nucleus in the expelled material. 
Future calculations with nuclear reaction networks will help to determine the
values more precisely.

\begin{figure}[ht]
\includegraphics[width=7.1cm,angle=-90]{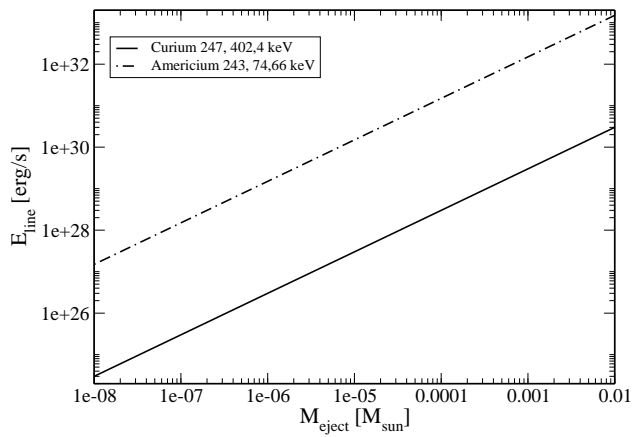}
\caption{Luminosity of two gamma ray lines as a function of the mass of
the specific isotope.}\label{trans}
\end{figure}

A long-term signature of the decay of radioactive nuclei might be 
observable in nearby remnants. A few of the radioactive actinides
are sources of gamma radiation. Hence gamma ray lines could be used to 
identify the
sites of past compact binary mergers. 
For example americium 243 with a half life time of 7370 years shows a strong
gamma ray line at 74,66 keV (Akovali \cite{akovali92}) and curium 247 with a 
half life time of $1.56\times
10^7$ years shows a strong gamma line at 402.4 keV (Schmorak \cite{schmorak92}).
For a given mass of the specific isotope (a few
percent of the ejecta, see above) a correlated 
luminosity in the corresponding gamma ray line will result 
(see Fig. \ref{trans}).
Gamma ray lines with energies above 511 keV are of
importance for observations as they lie above a possibly 
strong e$^+$e$^-$ annihilation background and 
other isotopes could be useful:
curium 248 will decay in spontaneous fission with the
emission of gamma ray photon lines  in the range of 94.9  to 605.91 keV
with a half-life time of $3.48\times10^5$ years. 
An example for a non actinide r-process
element  with certain interest for this problem is bismuth 208 with gamma 
emission in the 2.61 MeV line and a half life time of
$3.68\times10^5$ years (Martin \cite{martin86}).
The gamma ray satellite INTEGRAL is capable to observe $2 \times 10^{32}$ 
erg of emission in the 
americium line only in the very advantageous case of a remnant being at a 
distance of 1 kpc, assuming an exposure of 10$^6$ s.
But we note that most of the energy range of the gamma ray lines mentioned 
above is well suited to be 
observed with instruments using focusing optics for soft gamma rays by
Laue lenses, which will achieve a by orders of magnitude better sensitivity
(von Ballmoos \& Smither \cite{vonballmoos94}, Pisa et al. 
\cite{pisa04} and Frontera et al. \cite{frontera05}).
The strength of the total galactic gamma ray emission of heavy r-process 
elements might help to estimate the merger rate of compact binaries in the
galaxy.
Once remnants of compact binary mergers
are identified in this way, gamma ray lines might even be used to investigate
the chemical composition of the ejected material.

\subsection{Bremsstrahlung and X-ray line emission}\label{brems}

The thermal bremsstrahlung emission of the remnant
during the early evolution will be dominated by the ejecta. Since the mass of
the ejecta in compact binary mergers is much smaller than in supernova
explosions the density inside the remnants of these mergers will
also be much smaller than the density in SNRs at a comparable
state. Additionally the rarefraction of the interior of the exploding bubble
will happen much faster in merger sites as they expand with
very high velocities. 
Lower densities in the remnants will also have an impact on
the emissivity of the material. For example the bremsstrahlung 
emissivity of the 
remnant scales with $\rho^2\epsilon^{1/2}$, with $\rho$ being the 
density of the
medium and $\epsilon$ being the specific internal energy (see Ayal \& Piran
\cite{ayal01}). Hence due to the lower densities, the bremsstrahlung luminosity 
of remnants of 
mergers might be significantly fainter than SNRs with a
similar size. But several other processes will influence this result.
The ejecta of mergers may consist of very heavy r-process nuclei
(see also Sect. \ref{rproc}).
High abundances of heavy nuclei can enhance the thermal bremsstrahlung
(for a discussion on iron see Brighenti \& Mathews \cite{brighenti05}). 
Additionally very heavy elements are only fully ionized at very high 
temperatures. For example
the binding energy of the K shell in uranium is about 115.6 keV
(Bearden \& Burr \cite{bearden67}) which means
that uranium is fully ionized above a temperature of $1.3 \times 10^9$ K
and line emission of Uranium is possible at lower temperatures.
X-ray line emission of heavy elements 
may partly compensate the
effect of decreased bremsstrahlung emission due to the low densities inside
the remnants. The impact of these effects on the
emissivity of the interior of the expanding bubbles has to be further analyzed 
in future.

\subsection{Synchrotron emission}\label{sync}

The synchrotron emission during the early evolution of the remnant will be
mainly produced by the ejecta, which is exceptionally neutron rich and 
electron poor.
As the expelled material consists of very heavy r-process elements (see Sect.
\ref{rproc}),
the high electron binding energy of very neutron rich and
heavy nuclei will prevent the electrons to decouple from the ions (see Sect.
\ref{brems}).
This will reduce the number of free electrons which are available for
synchrotron radiation.
The lack of free electrons in the expanding medium will lead to   
decreased synchrotron emission.
It is further interesting to note that also the synchrotron
volume emissivity is quite a strong function of the density. 
It scales with $\rho^2\epsilon^{2}$ of the emitting material for appropriate 
assumptions (for more details see Ayal \& Piran 
\cite{ayal01}). Here again remnants of compact binary mergers will appear
much fainter than supernova remnants due to their low interior densities.

\subsection{Hard gamma ray emission}

SNRs are known to be 
sources of hard gamma rays in the TeV range (see Aharonian et al. 
\cite{aharonian05a}).  Hard gamma rays are produced in the interaction of
accelerated particles (for details see Aharonian et al. \cite{aharonian05a}).
We expect remnants of compact binary mergers also to radiate in hard gamma
rays similarly to SNRs. Hence remnants of mergers
may appear as sources of hard gamma rays, but due to the constraints 
of Sect. \ref{brems} and \ref{sync}, with a significantly reduced
bremsstrahlung and synchrotron radiation: So they might represent
a candidate for recently discovered 'dark accelerators' (Aharonian et al. 
\cite{aharonian05a},
\cite{aharonian05b}) but this has to be further investigated as well. 
It is also important to mention that at least in the case 
of HESS J1303-631 the hard gamma ray emission seems to be connected to an
association of young stars (see Aharonian et al. \cite{aharonian05b}) which
is difficult to explain with the proposed scenario.


\begin{acknowledgements}
We thank the referee for helpful comments.
We acknowledge the support of the European Commission through grant number
RII3-CT-2003-506079 (HPC-Europa) and FWF through P15868. We want to thank 
Marialuisa Aliotta, Roland Diehl, Thomas Janka, Alex Murphy, Sabine Schindler 
and Philip Woods for enlightening discussions.
\end{acknowledgements}



\begin{thebibliography}{}

\bibitem[2005a]{aharonian05a}
Aharonian, F., Akhperjanian, A. G., Aye, K.-M. et al. 2005, Science, 307, 1938

\bibitem[2005b]{aharonian05b}
Aharonian, F. Akhperjam, A. G., Aye, K.-M. et al. 2005, astro-ph/0505219

\bibitem[1992] {akovali92}
Akovali, Y. 1992, Nuclear Data Sheet 66,897

\bibitem[2001]{ayal01}
Ayal, S. \& Piran, T. 2001, ApJ, 555, 23

\bibitem[1967]{bearden67}
Bearden, J. A. \& Burr, A. F. 1967, Rev. Mod. Phys., 39, 125

\bibitem[2005]{bloom05}
Bloom, J. S., Prochaska, J. X., Pooley, D. et al. 2005, astro-ph/0505480

\bibitem[2005]{brighenti05}
Brighenti, F. \& Mathews, W. G. 2005, ApJ in press, astro-ph/0505527

\bibitem[1974]{chevalier74}
Chevalier, R. A. 1974, ApJ, 192, 457

\bibitem[1977]{chevalier77}
Chevalier, R. A. 1977, ARA\&A, 15, 175

\bibitem[1996]{dorfi96}
Dorfi, E. A. \& V\"olk, H. J. 1996, A\&A, 307, 715

\bibitem[1998]{efremov98}
Efremov, Y. N., Elmgreen, B. G. \& Hodge, P. W. 1998, ApJ, 501, L163

\bibitem[1999]{freiburghaus99}
Freiburghaus, C., Rosswog, S. \& Thielemann, F.-K. 1999, ApJ, 525, 121

\bibitem[2005]{frontera05}
Frontera, F., Pisa, A., De Chiara, P. et al. 2005, Proc. of the 39th ESLAB
Symposium, astro-ph/0507175

\bibitem[2005]{gehrels05}
Gehrels, N., Barbier, L., Bathelmy, S. D. et al. 2005, astro-ph/0505630

\bibitem[2004]{goriely04}
Goriely, S, Demetriou, P., Janka, H.-T., Pearson, J. M. \& Samyn, M. 2004,
Nucl. Phys. A, astro-ph/0410429

\bibitem[2005]{hatzidimitrou05}
Hatzidimitrou, D., Stanimirovic, S., Maragoudaki, F. et al. 2005, 
MNRAS, 360, 117

\bibitem[1993]{kouveliotou93}
Kouveliotou, C., Meegan, C., Fishman, G. J. et al. 1993, ApJ, 413, L101

\bibitem[1974]{lattimer74}
Lattimer, J. M. \& Schramm, D. N. 1974, ApJ, 192, L145

\bibitem[1999]{lee99}
Lee, W. H., Klu\'zniak, W. \& Lodzimierz, W. 1999, 526, 178

\bibitem[1998]{loeb98}
Loeb, A. \& Perna, R. 1998, ApJ, 503, L35

\bibitem[1998]{li98}
Li, L.-X. \& Paczy\'nski, B. 1998, ApJ, 507, L59

\bibitem[1986]{martin86}
Martin, M. J. 1986, Nuclear Data Sheet 47,797

\bibitem[1986]{paczynski86}
Paczy\'nski, B. 1986, ApJ, 308, L43

\bibitem[2000]{perna00}
Perna, R., Raymond, J. 2000, ApJ, 539, 706

\bibitem[2004]{pisa04}
Pisa, A., Fontera, F., De Chiara, P. et al. 2004, SPIE Proc., 5536, 39
astro-ph/0411574

\bibitem[1999]{rhode99}
Rhode, K. L., Salzer, J. J., Westpfahl, D. J. \& Radice, L. A. 1999, AJ, 118,
323

\bibitem[2002]{rosswog02}
Rosswog, S. \& Davies, M. 2002, MNRAS, 334, 481

\bibitem[2005]{rosswog05}
Rosswog, S. 2005, ApJ in press, astro-ph/0508138

\bibitem[1996]{ruffert96}
Ruffert, M., Janka, H.-T. \& Schaefer, G. 1996, A\&A, 311, 532

\bibitem[1997]{ruffert97}
Ruffert, M., Janka, H.-T., Takahashi, K. \& Sch\"afer, G. 1997, A\&A, 319, 122

\bibitem[1992]{schmorak92}
Schmorak, M. R. 1992, Nuclear Data Sheet 66,839

\bibitem[2005]{shibata05}
Shibata, M., Taniguchi, K. \& Uryu, K. 2005, Phys. Rev. D71 084021

\bibitem[2005]{simpson05}
Simpson, C. E., Hunter, D. A. \& Knezek, P. M. 2005, AJ, 129, 160

\bibitem[2004]{stairs04}
Stairs, I. H. 2004, Science, 304, 547

\bibitem[2005]{tang05}
Tang, S., Wang, Q. D. 2005, ApJ, 628, 205

\bibitem[1994]{vonballmoos94}
von Ballmoos, P., Smither, R. K. 1994, ApJS, 92, 663

\end{thebibliography}
\end{document}